\documentclass{aa} \usepackage{graphicx}\usepackage{rotating}

\def\simlt{$\; \buildrel < \over \sim \;$}
\def\ltsima{\lower.5ex\hbox{\simlt}}
\def\simgt{$\; \buildrel > \over \sim \;$}

\def\gtsima{\lower.5ex\hbox{\simgt}}
\def\kms{km s$^{-1}$}

\def\etal{{et~al.~}}

\def\xmm{{\it XMM-Newton~}}
\def\sax{{\it BeppoSAX~}}
\def\chandra{{\it Chandra~}}

\def\cm2{\rm{cm}^{2}}
\def\cmM2{\rm cm$^{-2}$}

\def\Ms{M_\odot}

\def\csq{\chi^2}

\def\le{{_<\atop^{-}}}

\def\41{{NGC~4151}}

\begin{document}

\title{Clearing up the clouds around NGC~4151: evidence of a highly ionized absorber}

\author{ L. Piro \inst{1} \and  A. De Rosa  \inst{1} \and G. Matt
  \inst{2} \and G. C. Perola  \inst{2}}

\institute{ {Istituto di Astrofisica Spaziale e Fisica Cosmica,
INAF, Via Fosso del Cavaliere, 00133 Roma, Italy}\and
{Dipartimento di Fisica, Universit\`a degli Studi ``Roma Tre'',
Via della Vasca Navale 84, I--00146 Roma, Italy}}

\offprints{Luigi Piro: piro@rm.iasf.cnr.it}
\date{ }
\titlerunning{Outflow in NGC 4151}

\abstract{The Seyfert 1 galaxy \41 is characterized by complex
X-ray absorption, well described by a dual absorber, composed of a
uniform mildly ionized gas and a cold system that partially covers
the central source. However,  in one of the 5 \sax observations,
the spectrum shows two peculiar features. An absorption feature is
detected around 8.5-9 keV with a statistical significance of
99.96\%. This feature can be fitted either with an absorption edge
at E$=8.62^{+0.34}_{-0.52}$keV with optical depth
$\tau=0.06\pm0.03$ or with an absorption line with
 $9.5^{+1.3}_{-0.6}$ keV,
width $\sigma=0.95^{+1.2}_{-0.7}$ keV and EW= 200 eV. In the first
case, we associate the feature to highly ionized iron at rest,
like FeXXII-FeXXIII (${\rm E_{rest}}$=8.4-8.5 keV). In the second
case the feature could be identified with a blend of FeXXV and
FeXXVI lines, with an outflow velocity v$ \approx (0.09-0.26)c$.
This spectrum is also characterized by a substantial reduction of
the absorption column density and the covering fraction of the
dual absorber.  In particular the column density of the mildly
ionized and cold absorbers is $\approx 3-5$ times lower than
observed in the other states, and the covering fraction is reduced
by $\approx$ 40 per cent. We propose a possible explanation
linking the two properties in terms of a multi-phase ionized
absorber.
}

\maketitle

\section{Introduction}
\label{intro}

\chandra and \xmm observations have shown evidence of narrow
absorption lines (NAL) in X-ray spectra of Seyfert 1 galaxies
(e.g. NGC~3783 \cite{kaspi02}). The lines are often blueshifted,
suggesting that the material where the NAL originate is flowing at
velocities of a few hundred \kms. This is the same medium that
produces absorption edges from ionized species (warm absorbers)
but that is usually transparent in the Fe K spectral band above 6
keV. Gas outflowing at higher velocities has been recently found
in the spectra of several quasars (PG1211+143, \cite{pou03a};
PG0844+349 \cite{pou03b}; PDS 456, \cite{reeves03},
APM~08279+5255, \cite{chartas02}), through blueshifted absorption
lines. In these high luminosity sources the wind was found at
velocity from 500 up to  120,000 \kms  (but see also
\cite{kaspi04,mckernan04}). In Seyfert galaxies no strong evidence
of outflows with such large velocities has been found so far (but
see the case of Mkn~509, \cite{dadina05}).

 In this letter we report on the peculiar
absorption properties observed in one of the five observations of
the bright Sy1 galaxy \41 by \sax. The broad-band analysis of
these observations can be found in \cite{piro05}  (hereafter P05).
The absorber in \41 is usually well described by a dual absorber
(PO5 and references therein): a cold one, associated with BLR
clouds, that partially covers the source, and an external uniform
screen, mildly photoionized by the primary continuum. In the
December 2001 spectrum, however, there is also evidence of an
absorption feature around 8-9 keV, along with a very low
absorption in both the cold and mildly ionized absorbers. In
particular, the column densities were lower by a factor of 3-5,
and the covering fraction of the cold gas was reduced by about 40
per cent. It is worth noting that this combination of low
absorption (both in column density and in covering fraction) has
never been observed in \41. In Section \ref{feature} we test
different spectral models to reproduce the absorption feature.
These models will be discussed in Section \ref{discussion} and our
conclusions will be drawn in Section \ref{conclusion}. We adopted
the standard  prescription (\cite{fiore99}) for  data reduction of
\sax observations (P05).   Hereafter errors and upper limits on
spectral parameters correspond to $\Delta \csq=2.7$, i.e. 90 per
cent confidence level for a single parameter of interest.

\section{Evidence of absorption feature}
\label{feature}

The December 2001 spectrum, with net exposure times of 114 ks in
the MECS(1.8-10 keV), 43 ks in the LECS(0.1-2 keV) and 53 ks in
the PDS(13-200 keV), was the longest taken by \sax.  The spectra
of the three instruments were simultaneously fitted with a {\it
baseline model} (BLM) which includes the following components:

A.  The primary continuum described  by ${\rm C E^{-\Gamma}
exp{(-E/E_c)}}$ i.e a power law with an exponential cut-off. C,
${\rm \Gamma}$ and ${\rm E_c}$ are free parameters.

B. A Compton reflection component from cold matter (PEXRAV model
in XSPEC, \cite{pexrav_ref}), with the cosine of the inclination
angle of the reflector fixed to $0.95$ and the spectral slope
linked to that of the primary continuum.  The normalization is a
free parameter.

C. A narrow K$\alpha$ iron line modelled with a gaussian profile
with the intrinsic width set to 10 eV,  with intensity and energy
as free parameters.

D. The soft X-ray spectrum (E $< 2$ keV) is fitted with a
combination of a thermal component model (MEKAL model in XSPEC
with the temperature fixed to 0.15 keV  and free normalization)
and a scattering component by ionized material (described by a
power--law with the same slope of the primary continuum  and a
free normalization)

E. The complex absorption is modelled with a dual absorber model.
A fraction ${\rm f_{cov}}$ of the source is covered by a cold
column density ${\rm N_{Hcov}}$. An additional uniform and ionized
gas is responsible for additional absorption of the spectrum at
low energies and for a mildly ionized Fe edge at $\sim$ 7.5 keV
(FeVII-FeVIII). This {\it mildly ionized} gas is characterized by
a column density ${\rm N_{Hwarm}}$ and a ionization parameter
${\rm \xi=L_{ion}/nR^2}$, where ${\rm L_{ion}}$ is the source
luminosity from 5 eV to 300 keV,  ${\rm n}$ is the hydrogen number
density of the gas (in \rm cm$^{-3}$) and R its distance from the
central ionizing source (in \rm cm) (model ABSORI in XSPEC,
\cite{done92}).  The parameters ${\rm f_{cov}}$, ${\rm N_{Hcov}}$,
${\rm \xi}$ are left free.

We assume element abundances from Anders \& Grevesse (1989). The
Fe abundance in the cold and mildly ionized absorbers and in the
reflection component is set to one and the same value, and left
free to vary during the fit. This model assumes that only the
cutoffed power-law and the cold reflection are subjected to
complex absorption. Additional absorption through our own Galaxy
is applied to {\it all} the emission components (${\rm
N_{Hgal}=2\times 10^{20}}$ \rm cm$^{-2}$).

\begin{figure}
\centering
\includegraphics[height=8cm,width=9cm,angle=0]{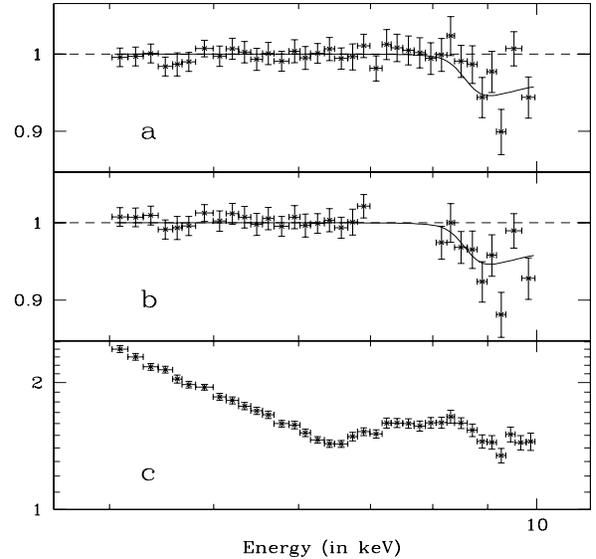}
\caption{A zoom in MECS 4-10 keV of data/model ratios when the
spectrum is fitted  with BLM (panel a) or with the simplified
model described in the text (panel b).  Data have been binned
adopting 3 bins per resolution element (FWHM). This choice
maximize the S/N ratio per bin still providing the minimum
over-sampling to retain the spectral information. The continuous
line is the profile of the absorption edge convoluted through the
MECS response. Panel c) shows the ratio of Dec.2001 to the sum of
the other spectra.} \label{datamodel}
\end{figure}

\begin{figure}
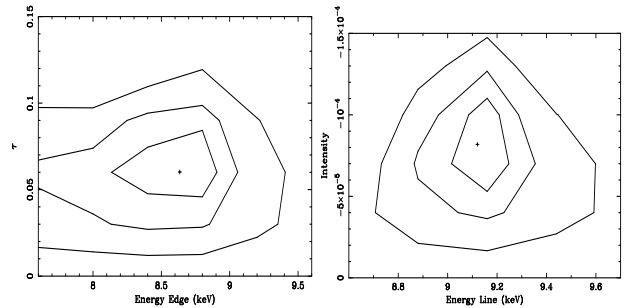

\centering
\includegraphics[height=4cm,width=4cm,angle=-90]{Hg041_f2a.ps}
\includegraphics[height=4cm,width=4cm,angle=-90]{Hg041_f2b.ps}
\caption{Contour plot of the edge optical depth vs. energy (left)
and absorption line intensity vs energy (right).  Confidence
levels are at 68\%,90\%,99\% } \label{contour}
\end{figure}

\begin{table}
\centering
\caption[] {Best fit parameters fitting the absorption feature in \sax
observation in Dec 2001}
\begin{flushleft}
\begin{tabular}{l c c c c}
\noalign {\medskip} \hline
 Model$^6$ &   $^1$E;$^2\xi$  &
$\tau$;$^3$N${\rm _{H,Fe}}$;$^4$EW &  z & $^5\chi^2$/dof \\
\hline
\noalign{\medskip}
Edge & $8.62^{+0.34}_{-0.52}$ & $0.06^{+0.03}_{-0.03}$ & - & 58/70\\
Ion Abs & $>$300 & $3.0^{+1.8}_{-1.8}$ &$<$0.07 & 63.2/69\\
Abs line & $9.1^{+0.2}_{-0.2}$ & $100^{+45}_{-45}$ & 0.09-0.26 & 57.9/70\\
\hline \noalign{\noindent Note: $^1$ In keV; $^2$ In \rm erg cm
s$^{-1}$; $^3$ N${\rm _{H,Fe}}$=N${\rm _H \times A_{Fe}}$, in
10$^{22}$\rm cm$^{-2}$; $^4$ In eV; $^5$ BLM:
$\chi^2$/dof=71.4/72; $^6$ Component added to BLM}
\end{tabular}
\end{flushleft}
\label{fit}
\end{table}

The residuals of the fit with the BLM to the spectrum of Dec 2001
in the  MECS  4-10 keV range (presented in Fig \ref{datamodel},
panel a) show evidence of an absorption feature around 9 keV. We
can exclude that this feature is an artifact of the background or
of the complex modelling required to describe the broad-band
spectrum of NGC~4151. The background spectrum in the MECS in the
8-10 keV range is flat and it is more than two orders of magnitude
lower than the source in the extracted source region. We note that
in fitting the BLM, we left free to vary all the parameters
specified above, thus tending to minimize deviations. We have
additionally tested the robustness of the result as follows. First
we have adopted a simpler model to fit the MECS and PDS data only.
To avoid the complexity of modelling with a dual cold and mildly
ionized absorber, we exclude the  data below 4 keV and in the 7-8
keV, and adopted a (cut-off) power law with a single cold absorber
plus reflection and a narrow emission line. All the parameters
were left free. The residuals, shown in panel b) of Fig
\ref{datamodel}, show a even more prominent absorption feature.
Finally, we employ a model-independent test, by producing the
ratio of  the spectrum of Dec.01 to the sum of spectra of all the
other observations (where we do not find evidence of absorption
feature, see below). This is shown in panel c) of Fig
\ref{datamodel}. Above 8.3 keV the ratio decreases sharply and is
significantly below  the adjacent points at lower energies,
supporting the robustness of the absorption feature. We note that
the presence of an iron line at 6.4 keV less variable than the
intrinsic continuum (that was brighter in Dec.01 compared to the
other observations) compresses the amplitude of variations in the
5-7 keV region. A discussion on this issue goes beyond the scope
of this paper and is reported elsewhere (P05). Here it is
important to underline another feature that characterizes the
spectrum of Dec.01 in comparison with the others. Below 5.5 keV
the ratio shows that the spectrum of Dec.01 is markedly softer. In
fact, the fit with BLM shows that this is entirely due to a
substantial decrease of the dual absorber in Dec.01
($f{\rm_{cov}=0.34\pm0.07, N_{H,cov}=(3.5\pm0.9)\times 10^{22}}$
\cmM2, ${\rm N_{H,warm}=(0.9\pm0.2)\times 10^{22}}$ \cmM2 vs ${\rm
f_{cov}=0.6\pm0.04, N_{H,cov}=(19\pm4)\times 10^{22}}$ \cmM2 ,
${\rm N_{H,warm}=(7\pm1)\times 10^{22}}$ \cmM2 derived from the
summed spectrum of the other observations).

 We attempted to
reproduce the absorption feature with different models; the best
fit values are shown in Table \ref{fit}. First we added to the
model an absorption edge, with energy E and optical depth $\tau$
as free parameters.  The improvement with respect to the BLM is
$\Delta \csq=13$. This corresponds to a confidence level of 99.98
\%  for the addition of two parameters, i.e. taking into account
that the energy was not known a priori. The energy of the edge was
found at E=$8.62^{+0.34}_{-0.52}$ keV and the optical depth is
$\tau=0.06^{+0.03}_{-0.03}$ (see the first line in Table \ref{fit}
 and the contour plot in Fig. \ref{contour}).  A similar
result is derived from the addition of the edge to the simplified
model described above. The chi square improves from
$\csq/\nu=64.4/37$ to $\csq/\nu=38.4/35$ upon the addition of the
edge with a slightly more significative confidence level of
99.99\%. The profile of the edge (convoluted through the MECS
response matrix), is shown in Fig.\ref{datamodel}.

To have a more physical representation of the Fe absorption we
fitted the \sax spectrum of Dec 2001 adding to our BLM a uniform
highly ionized gas (ABSORI in XSPEC, \cite{done92}). Iron
abundance in the gas is the same as in the cold and mildly ionized
absorbers and left free during the fit.
 This model gave us a good fit with
$\Delta \csq=8$ (with the addition of three free parameters), with
respect to the BLM (see the second line in Table \ref{fit}). The
gas is highly ionized ( $\xi \gtsima 300$ erg cm s$^{-1}$) and it
is characterized by a column density of ${\rm
N_{H,ion}=5^{+3}_{-3}\times 10^{21}}$ \cmM2 and iron abundance
${\rm A_{Fe}=6.1^{+1.9}_{-1.1}}$. We took into account a possible
blueshift and we found an upper limit v/c $<0.07$. A less ionized
medium is excluded because the column density inferred from the
edge depth (${\rm N_{H,Fe}=3\times 10^{22}}$ \cmM2) would
significantly affect the low energy spectrum. This in turn sets
the upper limit to the velocity found above.

Finally we fitted the feature with an absorption line with a
narrow gaussian profile. This model gave $\Delta \csq=13$ (with
the addition of two free parameters) with respect to the BLM (last
line in Table \ref{fit}). The energy of the line was found at
E$=9.1^{+0.2}_{-0.2}$ keV with an  equivalent width $100\pm45$ eV
(Fig. \ref{contour}). When left free to be broad, the line is
reproduced with energy $9.5^{+1.3}_{-0.6}$ keV, an intrinsic width
$1.0^{+1.15}_{-0.80}$ keV and EW$\sim$ 200 eV. However the fit is
not significantly better than that with a narrow line.

We do not find statistically significant evidence of a similar
feature in the other \sax observations . We have then checked if
we can {\it exclude} the presence of a feature with the same
properties as that observed in Dec 2001. This has been done by
adding an edge with the energy fixed at 8.6 keV. The resulting
upper limits - for each single observation - are above, i.e.
consistent, with the detection. This is not surprising,
considering the much higher statistics available in Dec 2001. A
tighter upper limit $\tau\le 0.04$ is derived from the summed
spectrum of the other observations, a value that is below but
still marginally consistent with the detection of Dec 2001.
Considering this test as another independent search on a spectrum
with a similar statistical weight as that of Dec.2001, we
conservatively adjust the significance of the feature to 99.96 \%.

We performed the same analysis on the \xmm spectra of \41
available in the public archive. \xmm observed \41 on 2000 Dec
21-22-23 and on 2003 May 25-26-27. The longer exposures were those
of 2000 Dec. 22 and 21 ($\sim$ 62 ks and $\sim$ 34 ks,
respectively), the other observations were about $\sim$ 20 ks
long. We fitted all the spectra with our BLM, and we looked for
the presence of an absorbing feature. We found only an upper limit
to the optical depth of an absorption edge with the energy fixed
to the value observed by \sax (8.62 keV) of $\tau\le 0.02$. It is
worth to note that none of these observations were characterized
by a very low absorption as detected in Dec. 2001 by \sax.

\section{Discussion}
\label{discussion}

Let us first discuss the interpretation of the feature with an
absorption line. Candidate resonant absorption lines are at 6.70
keV (FeXXV K$\alpha$), 7.88 keV (FeXXV K$\beta$) 6.97 keV (FeXXVI
K$\alpha$) and 8.25 keV (FeXXVI K$\beta$), depending on the
ionization state. Assuming a dispersion of $\sim$ 1000 \kms, the
measured equivalent width, $\sim$ 100 eV, would indicate a column
density of about N${\rm _H \sim 10^{23}}$ \cmM2 (\cite{bianchi05})
and a ratio K$\alpha$/K$\beta$ of about 1.6 and 2.6 for FeXXVI and
FeXXV respectively  (\cite{risaliti05}). Taking the weighted
average of the K$\alpha$ and K$\beta$ energies and comparing with
the observed energy, we derive an outflow velocity of v/c$ \sim$
0.09-0.26, depending on the association FeXXV or FeXXVI.

\41 would be  one of the first case of high velocity ionized
outflow through the detection of blueshifted Fe K absorption line
in the X-ray spectrum of a Sy 1 with low accretion rate. Evidence
of this feature has instead been found by \xmm and \chandra in
sources accreting close to the Eddington rate and requires
 an outflow of highly ionized gas at high
velocity and with a large column density (e.g.  PG1211+143 N${\rm
_H\sim 5\times 10^{23}}$ \cmM2 $\log{\xi}\sim$ 3.4 and v$\sim
24,000$ \kms; \cite{pou03a}, PG0844+349 N${\rm _H\sim 4\times
10^{23}}$ \cmM2 $\log{\xi}\sim$ 3.7 and v$\sim 60,000$ \kms,
\cite{pou03b}, PDS~456 N${\rm _H\sim 5\times 10^{23}}$ \cmM2
$\log{\xi}\sim$ 2.5 and v$\sim 50,000$ \kms, \cite{reeves03}).
These value are in good agreement with those found by \sax in the
case of \41.

However the outflow rate in  those quasars is comparable to the
mass accretion rate, suggesting that the outflow could originate
by wind driven by the radiation pressure of the accretion disk
around the super-massive black hole. The case of \41 looks
different. With ${\rm L_{Edd}=1.8\times 10^{45}}$  erg s$^{-1}$
and ${\rm L_{bol} \sim 10^{44}}$ erg s$^{-1}$, the accretion is
taking place in a sub-Eddington regime, and therefore radiation
pressure can not drive the outflow. In addition we can derive a
rough estimation of the mass outflow rate as follows. When
reproduced with an absorption line, the observed stage of
ionization, FeXXV or FeXXVI, indicates a ionization parameter
$\xi\gtsima 1000$ \rm erg cm s$^{-1}$. Given a ionization
luminosity of L${\rm _{ion}\sim 2\times 10^{43}}$ erg s$^{-1}$, an
upper limit to ${\rm nR^2\ltsima 2\times 10^{40}}$ \rm cm$^{-1}$
is obtained, where n is the number density of the gas and R its
distance from the Black Hole. The mass outflow rate is ${\rm
\dot{M}_{out}=\Omega nR^2m_p v}$ where ${\rm \Omega}$ is the solid
angle in steradian subtended by the outflow, ${\rm m_p}$ the mass
of a proton and v the outflow velocity. Then we obtain  ${\rm
\dot{M}_{out}\le 4 \Omega \Ms yr^{-1}}$. In a stationary regime
${\rm \dot{M}_{out}\sim \dot{M}_{infl} \sim \frac{L_{bol}}{\eta
c^2}\sim 0.02 \Ms yr^{-1}}$, with an efficiency $\eta$=0.1, thus
requiring a highly collimated outflow (${\rm \Omega\gtsima 0.005
 sr}$) directed in our line of sight, which is unlikely.

\begin{figure}
\centering
\includegraphics[height=8.cm,width=8.cm]{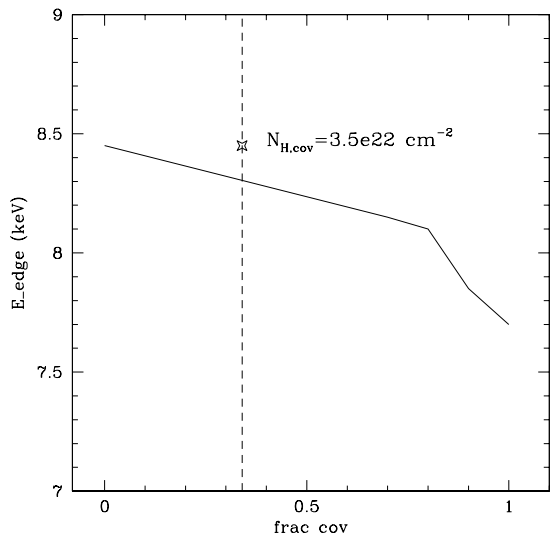}
\caption{XSTAR simulation. The energy of the Fe absorption edge in the
  external ionized gas is plotted versus the covered fraction of the
  internal cold gas with N${\rm _H=1.5\times 10^{23}}$ \cmM2. The case of
  Dec 2001 is plotted with a diamond. The incident continuum is
  a cut-off power--law with ${\rm \Gamma}$=1.7 and ${\rm E_c}$=130 keV. The plot
  can explain the appearance of the
  absorption  feature only in a peculiar state of \41 when the central
  source was almost uncovered.}
\label{ion_abund}
\end{figure}

Let us now discuss the association of the feature with an
absorption edge. In this case the energy of the edge is consistent
with a highly ionized gas with a low or zero velocity, so that the
problem of the mass outflow rate and its consistency with the
accretion flow does not apply. However this  excludes the
possibility to explain the transient presence of the feature in
terms of a short-lived outflowing phenomenon.\\
We have explored another solution, that takes into account the
coincidence between the appearance of the feature and the
uncovering of the central source. In this scenario the  gas
producing the absorption edge is located outside the cold
partially covering absorber. This gas is therefore subjected to a
different ionizing flux depending on the covering fraction
$f_{cov}$ and the absorption column density ${\rm N_{H,cov}}$.
This is shown in Figure \ref{ion_abund}, where the weighted
average energy of the absorption edge due to the most abundant ion
stages, computed with the XSTAR code (\cite{xstar}), is shown for
representative cases. We have adopted a power-law spectrum with
photon index ${\rm \Gamma}$=1.7 and a high energy cut-off at 130
keV as representative of the continuum. The intrinsic luminosity
of the source has been set to the value, ${\rm
L_{1Ry-1000Ry}=10^{43}}$ erg s$^{-1}$, as observed in Dec. 2001,
to check whether different stages of ionization can be obtained by
changing partial covering parameters only. The parameters of the
ionized absorber (i.e. ${\rm nR^2}$), were set by imposing that
the average energy of the edge for the spectrum observed in Dec.
2001 (${\rm f_{cov}=0.34}$ and ${\rm N_{H,cov}=3.5\times 10^{22}}$
\cmM2) is E$\sim$8.5 keV. The continuous line gives the result in
the case of ${\rm N_{H,cov}=1.5\times 10^{23}}$ \cmM2, i.e. the
typical value observed in \41, as function of ${\rm f_{cov}}$,
while the diamond corresponds to the case of Dec 2001. It is clear
that, when the covering fraction is $\gtsima 0.7$, as typically
observed in this source, the energy of the edge decreases to
around 7.6-8
keV,  i.e. near to that of the mildly ionized absorber.\\
In fact, it is tempting to identify the high and mildly ionized
absorber with the same system, i.e.  a multi-phase warm absorber,
in which the predominant ionization stage is determined by the
ionizing flux impinging on it. Interestingly, the column density
of the mildly ionized phase in Dec 2001 is the lowest ever
observed in \41, this  could (at least in part) due to the
transformation into the highly ionized phase. The sum of the mild
and high ionized column densities in Dec 2001 is $\sim
1.5\times10^{22}$ \cmM2. As mentioned above, in the other \sax
observations the gas should be mostly in a  mildly ionized phase,
so one would expect to observe a similar column density.
 However, the presence of variability, from 2 up to 8 $\times10^{22}$ \cmM2
(P05), does not allow an unambiguous conclusion.

We can derive an upper limit to the distance of the gas R to the
central source as follows.  By taking $\xi\gtsima 300$ \rm erg cm
s$^{-1}$ (see Table \ref{fit}), L${\rm _{ion}\sim 2\times
10^{43}}$ erg s$^{-1}$  and the column density $ 5\times 10^{21}$
\cmM2, we derive ${\rm R\ltsima 10^{19} f_V}$ cm, where ${\rm
f_V}$ is the volume filling factor of the gas. This value is fully
consistent with the variability of the observed absorption feature
on yearly time scales.

\section{Conclusions}
 \label{conclusion}

We presented a study of a peculiar state of the Seyfert 1 \41
observed by \sax on December 2001. The evidence in the spectrum of
an absorption feature around 9 keV suggests the presence of a
highly ionized gas. In the same spectrum there was a a substantial
decrease of the warm (mildly ionized) and cold absorbing gas that
characterize the complex absorption in \41. Both the column
densities and the cold covering fraction in Dec 2001 were lower
than in the other observations of the source. This combination has
never been observed before in \41.

A first possible scenario is one in which the absorbing feature is
identified with a blend of FeXXV and FeXXVI absorption lines,
produced in a gas with an outflow velocity v$ \approx
(0.09-0.26)$c , depending on the line identification.   However
the mass outflow rate is much higher than the mass accretion rate,
differently from the cases observed in high accretion rate
quasars, where similar lines were detected.

In a second and most appealing scenario the absorbing feature is
interpreted as an absorption edge of highly ionized iron at rest,
like FeXXII or FeXXIII (${\rm E_{rest}}$=8.4-8.5 keV). The gas
where this feature is produced is external to the cold partial
absorber, and it is therefore subjected to a different ionizing
flux depending on the covering fraction and column density of the
latter. This is fully consistent with the appearance of the
absorption feature only in a peculiar state of \41 when the
central source was almost uncovered.

It is then natural to identify the high and mildly ionized
absorber with the same system, i.e.  a multi-phase warm absorber, in
which the predominant ionization stage is determined by the fraction
of the continuum flux emerging from the inner cold absorber that
partially covers the central source.

{}

\end{document}